\documentclass[twocolumn,pre,showpacs,superscriptaddress]{revtex4}
\usepackage[utf8x]{inputenc}
\usepackage{epsfig}
\usepackage{amsmath}
\usepackage{comment}
\usepackage{color}
\usepackage[toc,page]{appendix}
\newcommand{\staticStructure}{S}

\newcommand{\citem}[1]{\color{red}[~]\color{black}}

\begin{document}

\title{From the granular Leidenfrost state to buoyancy-driven convection}

\author{Nicolas Rivas}
\affiliation{Multi-Scale Mechanics (MSM), MESA+, CTW, University of Twente, PO Box 217, 7500 AE Enschede, The Netherlands}
\author{Anthony R. Thornton}
\affiliation{Multi-Scale Mechanics (MSM), MESA+, CTW, University of Twente, PO Box 217, 7500 AE Enschede, The Netherlands}
\affiliation{Mathematics of Computational Science (MaCS), MESA+, CTW, University of Twente, PO Box 217, 7500 AE Enschede, The Netherlands}
\author{Stefan Luding}
\affiliation{Multi-Scale Mechanics (MSM), MESA+, CTW, University of Twente, PO Box 217, 7500 AE Enschede, The Netherlands}
\author{Devaraj van der Meer}
\affiliation{Physics of Fluids, University of Twente, The Netherlands}

\pacs{45.70.Qj, 46.65.+g, 05.40.-a}

\begin{abstract}
Grains inside a vertically vibrated box undergo a transition from a density inverted and horizontally homogeneous state, referred to as the granular Leidenfrost state, to a buoyancy-driven convective state. We perform a simulational study of the precursors of such a transition, and quantify their dynamics as the bed of grains is progressively fluidized. The transition is preceded by transient convective states, which increase their correlation time as the transition point is approached. Increasingly correlated convective flows lead to density fluctuations, as quantified by the structure factor, that also shows critical behaviour near the transition point. The amplitude of the modulations in the vertical velocity field are seen to be best described by a quintic supercritical amplitude equation with an additive noise term. The validity of such an amplitude equation, and previously observed collective semi-periodic oscillations of the bed of grains, suggests a new interpretation of the transition analogous to a coupled chain of vertically vibrated damped oscillators. Increasing the size of the container shows metastability of convective states, as well as an overall invariant critical behaviour close to the transition.
\end{abstract}

\maketitle

\section{Introduction}

Granular materials, defined as collections of dissipative particles large enough so that thermal fluctuations can be ignored, are an archetypal study case of complex dynamical systems. Decades of research have 
revealed many novel non-equilibrium phase transitions and collective 
behaviors~\cite{douady_subharmonic_1989,olafsen_clustering_1998,
clerc_liquidsolid-like_2008,ristow_pattern_2000,ottino_mixing_2000, 
aranson_patterns_2006}, the study of which not only has a fundamental physical 
interest, but is also relevant for many 
industries~\cite{muzzio_powder_2002,coussot_rheometry_2005,antony_granular_2004}. Many of these behaviors show a striking similarity with molecular fluids or 
solid phenomena~\cite{ramirez_thermal_2000, melo_transition_1994, 
thoroddsen_granular_2001, royer_high-speed_2009}, and some have even been 
successfully described by equilibrium 
theories~\cite{olafsen_two-dimensional_2005}. Studying the origin of these 
agreements advances our understanding of far-from-equilibrium states, and 
explores the limits of continuum descriptions of discrete systems. Furthermore, 
the low number of constituents, when compared to molecular counterparts, makes 
granular materials particularly suited for the study of noise effects in 
spatially extended transitions, a subject of increasing physical interest due to 
the ubiquitous presence of fluctuations in natural 
phenomena~\cite{goldman_noise_2004, clerc_liquidsolid-like_2008, 
ortega_subharmonic_2010, agez_bifurcations_2013, garcia-ojalvo_noise_1999}. 

In order to keep granular media fluidized it is necessary to provide energy to 
the system. Previously this has been  done in several distinct ways, as for 
example electromagnetically \cite{blair_clustering_2003, 
stambaugh_segregation_2004}, by shearing~\cite{miller_stress_1996}, or by 
boundary forces such as rotating a drum~\cite{seiden_complexity_2011} or 
vibrating the grains' container~\cite{wassgren_vertical_1996}. In vertically 
vibrated systems several complex collective dynamic behaviors have been 
observed, such as segregation~\cite{rosato_perspective_2002, 
windows-yule_inelasticity-induced_2014}, pattern 
formation~\cite{aranson_patterns_2006} and phase 
separation~\cite{olafsen_clustering_1998}. One particular case of the latter is 
the granular Leidenfrost state, where a dense, solid- or fluid-like region is 
sustained by a highly agitated low density gaseous region in contact with the 
vibrated bottom wall~\cite{meerson_close-packed_2003,eshuis_granular_2005}. It 
is so called because of the clear analogy with the water-over-vapor phenomenon 
observed in molecular fluids in contact with a high temperature 
surface~\cite{leidenfrost_aquae_1756}. If the vibration strength is increased, 
the Leidenfrost state evolves to a buoyancy-driven convective 
state~\cite{eshuis_onset_2010}, in analogy to Rayleigh-Bernard convection. 

In the following work we study the precursors of the transition from the 
granular Leidenfrost to the buoyancy-driven convective state in the context of 
bifurcations and critical theory. Previous experimental and simulational works determined the transition points as a function of the energy injection and the amount of particles in the system~\cite{eshuis_granular_2005, eshuis_onset_2010}. It was also shown that granular hydrodynamics is able to quantitatively capture the critical points of this instability, by performing a linear  stability analysis of perturbations over the Leidenfrost state~\cite{eshuis_onset_2010,eshuis_buoyancy_2013}. Here we explore further the regions close to the transition, motivated by the presence of complex transient dynamics which are expected to be dominated by fluctuations. This transition is an excellent candidate for studying the influence of fluctuations in hydrodynamic-like instabilities, due in part to its similarity with the Rayleigh-Benard instability present in regular fluids. Our goal is to increase the knowledge about the origin and evolution of the perturbations that lead to the instability, from both the microscopic and macroscopic perspectives, and relate the transition to other analogous dynamics through an unstable-mode amplitude equation.

After specifying the 
system and simulation methods (Sec. \ref{sec:System}), we begin by 
characterizing the two states involved in the transition (Sec. 
\ref{sec:Description}), and determining the phase space of the system by means 
of a convection intensity order parameter (Sec. \ref{sec:ConvectionIntensity}). 
With this, we are then able to study time-dependent transient convective states, 
that are present far below the transition and show a critically increasing 
correlation time as the transition is approached (Sec. 
\ref{sec:ConvectionTransient}). Furthermore, the static structure function 
allows us to study the evolution of the relevant length-scale in this pattern 
formation scenario, and see its behaviour prior to the transition (Sec. 
\ref{sec:StaticStructure}). Finally, it is observed that the amplitude of the 
critical pattern follows a growth ratio that is consistent with a quintic 
supercritical bifurcation, associated with parametrically driven spatially 
extended systems~ \cite{agez_bifurcations_2013, coullet_dispersion-induced_1994, 
clerc_localized_2008} (Sec. \ref{sec:Amplitude}). We suggest that the agreement 
with this universality class comes from the presence of collective semi-periodic 
oscillations, so called low-frequency oscillations (LFOs), present in density 
inverted systems \cite{rivas_low-frequency_2013}. All results are presented for 
different boundary conditions and sizes of the container, allowing us to observe 
the influence of confinement and variations of the total number of particles.

\section{System and simulations}
\label{sec:System}

\begin{figure}
  \begin{center}
  \includegraphics[scale=0.74]{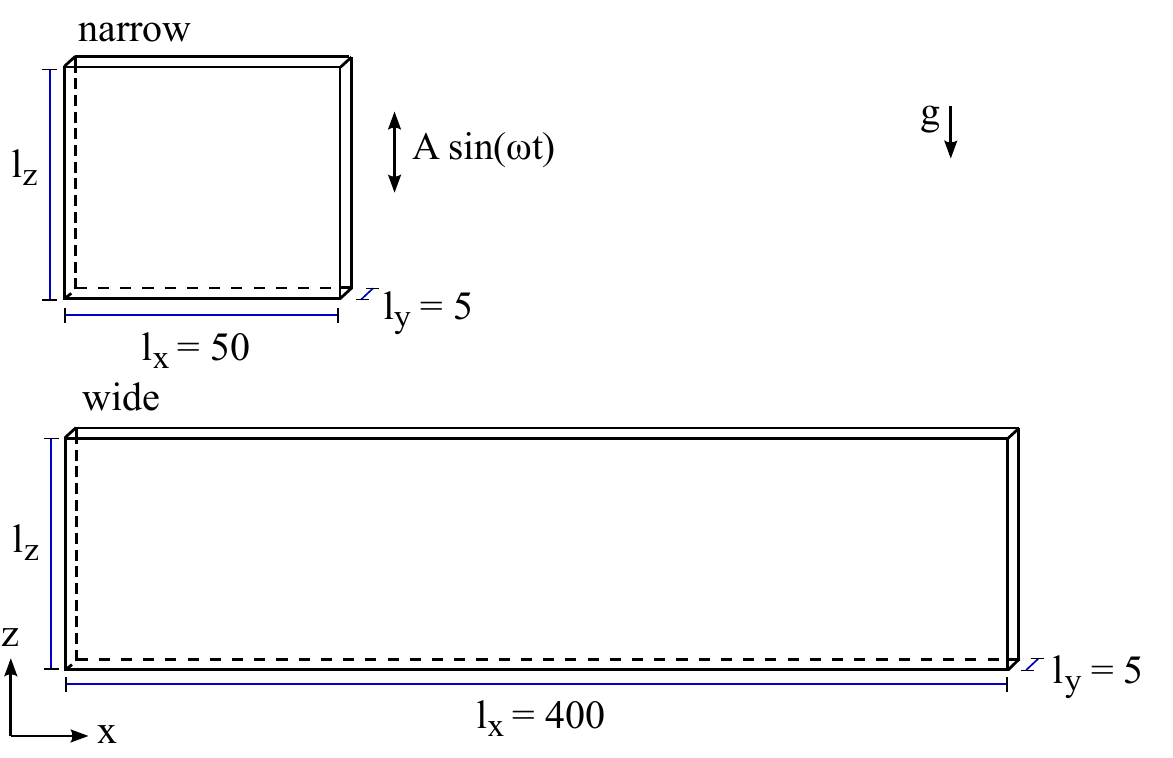}
  \end{center}
  \caption{
	    Schematic representation (not to scale) of the setup. Two different geometries are considered: narrow (top) and wide (bottom). Lengths are given in units of particle diameters $d$.
	  }
  \label{fig:Setup}
 \end{figure}

The setup consists of a quasi-two-dimensional rectangular box with open top, 
vibrated in the vertical direction. Two different box widths are considered, 
defining the \textit{narrow} system, with $l_x = 50$, and the \textit{wide} one, 
with $l_x = 400$. The depth of the container, on the other hand, is kept 
constant, $l_y = 5$; a schematic representation of the studied geometries is 
shown in Fig.~\ref{fig:Setup}. Here, and in what follows, we use dimensionless 
quantities with $d$ as lengthscale and $\sqrt{d/g}$ as timescale, and thus 
$\sqrt{g/d}$ as velocity units; when necessary, dimensional quantities will be 
distinguished by a tilde, i.e. $\tilde l_x = l_x d$. Grains are considered to be 
perfectly spherical, frictionless and monodisperse in size and mass. Their total 
number $N$ is determined by the number of filling layers $F \equiv N/(l_xl_y)$, 
which we fix at $F = 12$. Previous studies show that both the Leidenfrost and 
the buoyancy-driven convective states are observable for this number of 
layers~\cite{eshuis_phase_2007}. The whole box (base and side walls) is 
vertically vibrated in a bi-parabolic, quasi-sinusoidal way with a given 
frequency $\omega$ and amplitude $A$. The use of a quadratic interpolation 
instead of a sine function gives a considerable speed advantage in simulations, 
as the collision times with the moving walls can be predicted analytically. 
Previously, test simulations have been done using a sine function for exemplary 
cases, and no significant difference was 
observed~\cite{rivas_low-frequency_2013, mcnamara_energy_1998}. The amplitude of 
oscillation is kept fixed, $A = 0.1$, and thus the energy injection is 
controlled by the angular frequency $\omega$. The low amplitude is chosen to 
reduce as much as possible geometrical effects of the moving boundary (such as 
shock waves \cite{goldshtein_mechanics_1995}), and approximate the limit of a 
temperature boundary condition~\cite{soto_granular_2004}. Moreover, low 
amplitudes eliminate other inhomogeneous states for lower energies, such as 
undulations~\cite{eshuis_phase_2007}, which are not the object of this study. 
Overall, our selection of parameters is based on previous experimental 
setups 
where the transition was previously reported~\cite{eshuis_phase_2007,eshuis_onset_2010}.

The system is simulated using an event-driven (ED) hard-sphere algorithm. The 
advantage of using ED simulations over regular time-stepping methods is 
straightforward: computational speed. Even though the number of particles is 
relatively low ($\sim 10^4$), the high frequencies and very long physical 
times, of the order of hours, make the use of discrete particle methods (DPM) infeasible. In DPM 
simulations time-steps are constant and should be at least one order of 
magnitude lower than the collision time, which in itself must be at least an order of magnitude smaller than the lowest relevant time-scale, in our case $T = 2 \pi / \omega$ 
\cite{herrmann_modeling_1998}. Thus, for the high frequencies considered in our 
study, the small time-step prohibits to simulate in a practical time the long 
transients involved near a transition. On the contrary, the average time-step in 
ED is determined mainly by the density of the system, and not directly dependent 
on the frequency of oscillation of the container.

Collisions between particles are modeled by a normal restitution coefficient, 
$r_p = 0.9$~\cite{luding_granular_1995}. In order to avoid inelastic collapse, 
the TC model is used, where particle collisions are considered elastic if they 
occur within a given time, which we take as $t_c = 
10^{-5}$~\cite{luding_how_1998}. This essentially sets a lower limit for 
physically relevant velocities, as also slightly decreases the packing fraction 
of high density regions; possible relevant effects will be noted when 
appropriate.

Regarding boundary conditions, we consider both cases of periodic (PBC) and 
solid boundary conditions, with either elastic or dissipative walls (EBC and 
DBC, respectively). The different boundary types are only applied in the 
$x$-direction, as we would like to investigate the effects they have on the 
transition independent of other factors, as increased overall dissipation or 
free-volume; setting dissipative or periodic boundaries also in the 
$y$-direction would make the comparison less straightforward. Dissipative walls 
are set with the same restitution coefficient as between particles, $r_w = 0.9$. 
The effects of dissipative walls on convective states have already been studied 
in similar setups, both experimentally and 
numerically~\cite{windows-yule_thermal_2013}. Here we are interested in the 
effects of walls on the excitation or suppression of the modes relevant in the 
transition. Elastic walls (EBC) are used in order to see the influence of 
excluded volume effects near the sidewalls when comparing with periodic walls 
(PBC), as also to facilitate the analysis of fluctuations by fixing a reference 
frame. Furthermore, PBCs are used to study the dynamics of the bed of grains 
without confinement.

\section{Results}

\subsection{Macroscopic description}
\label{sec:Description}

\begin{figure}
  \begin{center}
  \includegraphics[scale=0.67]{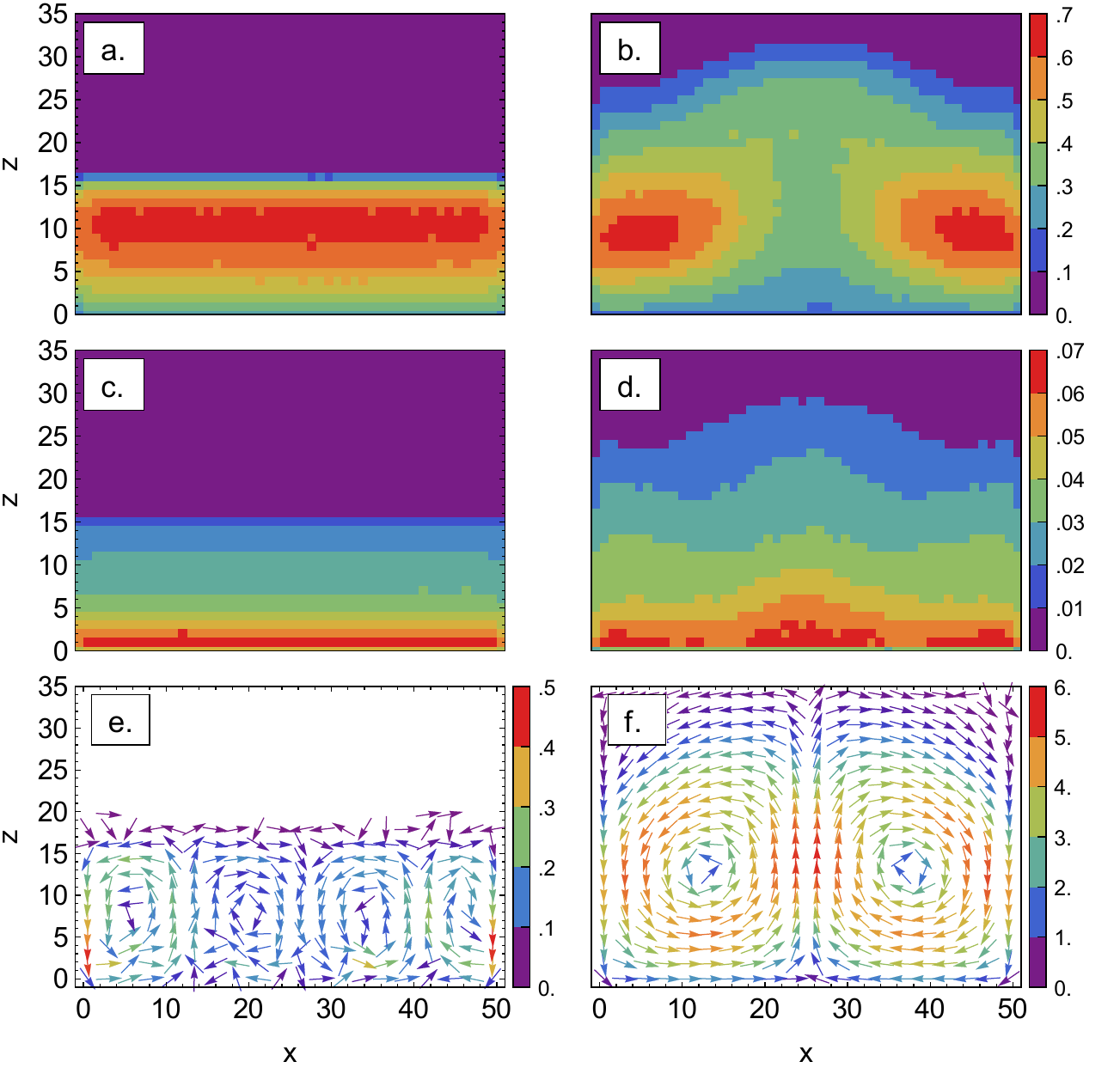}
  \end{center}
  \caption{
	   In the narrow system, time averaged number density of particles 
$\left<n(x,z)\right>_t$~(top), granular temperature $\left<T(x,z)\right>_t$ 
(middle) and velocity field $\left<\vec v(x,z)\right>_t$ (bottom), for systems 
in the granular Leidenfrost state (left) and in the buoyancy-driven convective 
state (right).
	  }
  \label{fig:2dfields}
 \end{figure}

The most evident difference between the granular Leidenfrost and buoyancy-driven convective states is the level of horizontal homogeneity. Fig.~\ref{fig:2dfields} shows time averaged number density $\left<n(x,z)\right>_t$, granular temperature $\left<T(x,z)\right>_t$ and velocity fields $\left<\vec v(x,z)\right>_t$ in each state, for narrow systems with EBC. The fields are obtained by binning the system in squares of size $d$. Time averages, $\left<\right>_t$, are always taken for at least $10^5\sqrt{d/g}$, which in dimensional terms for $d = 1$mm would correspond to experiments of about fifteen minutes. The granular temperature is defined as the kinetic energy of the fluctuating velocity, $3 k_B T \equiv m (\left<v^2\right>-\left<v\right>^2)$. The fields clearly show that in the Leidenfrost state, both $\rho$ and $T$ are homogeneous in the $x$-direction, while in the convective state the profiles are modulated by a dominant mode $k_c$. That is, the transition is \textit{morphogenetic}~\cite{prigogine_self-organization_1977}, as a pattern or new relevant length-scale arises from a homogeneous state. The convective mode defines the typical size of a pair of convective cells, $\lambda_c$; in the case shown in Fig.~\ref{fig:2dfields}, $\left<\lambda_c\right>_t \approx 50$, that is, $k_c = 1/\lambda_c \approx 0.02$.

It is important to remark that the buoyancy-driven convective state is also density inverted (see Fig.~\ref{fig:2dfields}b), and thus this characteristic is not a sufficient condition to define the Leidenfrost state. We demand two further properties for the system to be considered in this state: (a) higher density regions present distinct dynamics to the lower density ones (gas/fluid or gas/solid), to distinguish it from completely gaseous states~\cite{lan_macroscopic_1995}; and (b) the system remains horizontally homogeneous, to differentiate it from the convective state. In short, we define the granular Leidenfrost state as a density inverted, phase coexisting, horizontally homogeneous state.

As the energy input increases, the bed of grains in the dense region 
progressively looses its horizontal homogeneity, giving rise to convection; this 
is what we refer to as the granular Leidenfrost to buoyancy-driven convection 
transition or, in short, the LBC transition. In the following we define an order 
parameter based on the evolution of the velocity field, and observe its 
behaviour through the transition.

\subsection{Convection intensity}
\label{sec:ConvectionIntensity}

\begin{figure}
 \begin{center}
  \includegraphics[scale=0.70]{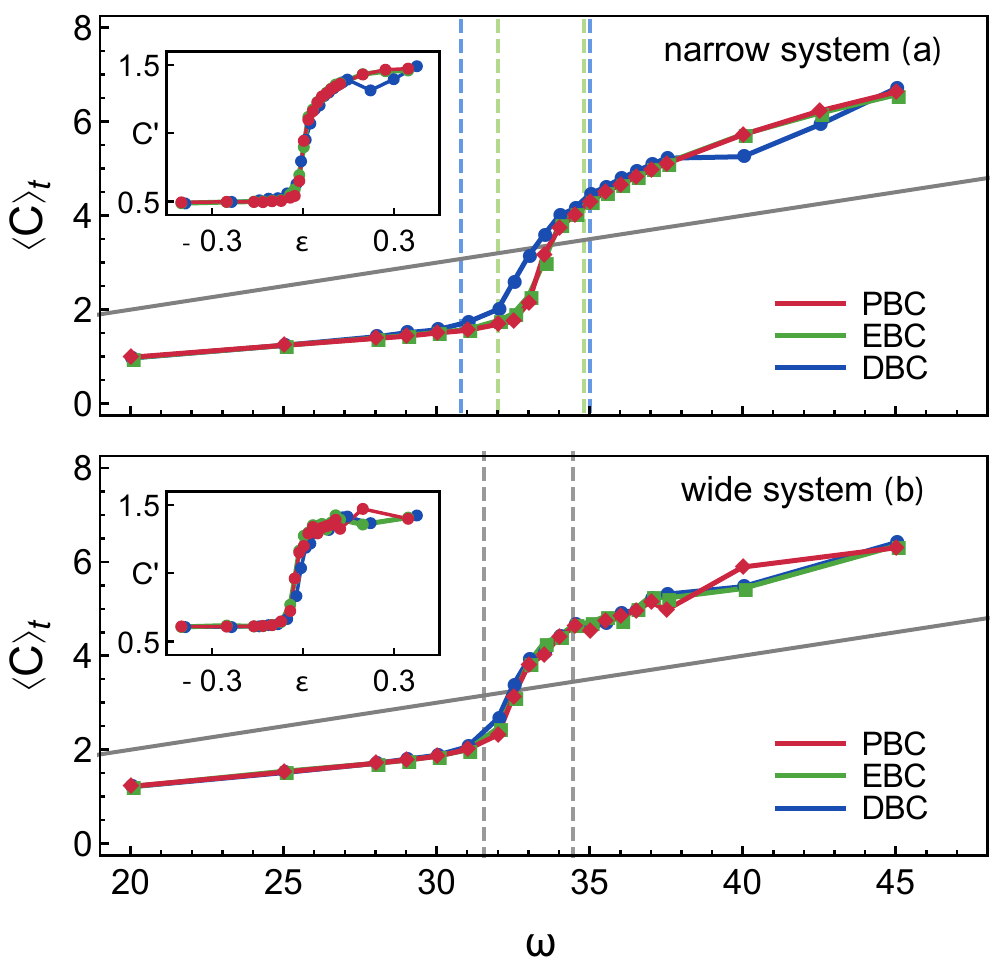}
 \end{center}
 \caption{Time averaged convection intensity $\left<C\right>_t$, defined in the main text \eqref{eq:convection_intensity}, as a function of the angular driving frequency $\omega$, for the narrow (top) and wide (bottom) containers with the boundary conditions indicated in the labels. Dashed lines indicate the transition region for the corresponding BC, as specified in the main text. The thick gray line corresponds to $A\omega$, the characteristic shaking velocity. The insets show the convection intensity normalized by the driving frequency, $C^* \equiv \left<C\right>_t/ A \omega$, as a function of the bifurcation parameter $\varepsilon = (\omega - \omega_c) / \omega_c$.}
 \label{fig:ConvectionStrength}
\end{figure}

For the study of critical behaviors it is of fundamental importance that the transition region between the two states is accurately measured. The different states can be easily distinguished by looking at the time-average velocity fields, which suggest the use of the convection intensity order parameter, defined as
\begin{equation}
  \label{eq:convection_intensity}
  C \equiv \tfrac{1}{2} \text{max}_z(\text{max}_x(v_z(x,z)) - \text{min}_x(v_z(x,z))).
\end{equation}
Here $v_z(x,z)$ is the scalar field of velocities in the $z$-direction, and the maxima are taken first over $z$ and then over $x$. In words, $C$ corresponds to half the highest difference of the vertical velocities at a particular height of the container. In a convective state $C$ is expected to be significantly higher than in a random flux case, due to the presence of stable upwards and downwards flux regions (as can be seen in Fig.~\ref{fig:2dfields}f). Even though the average vertical velocity is expected to scale with $A\omega$, the localization of the energy fluxes in the convective states is what produces a higher deviation, and thus a higher $C$. The time averaged convection intensity, $\left<C\right>_t$, captures the transition as a rapid increase with $\omega$, as shown in Fig.~\ref{fig:ConvectionStrength} for all considered systems.

In the Leidenfrost state $\left<C\right>_t$ increases linearly with $\omega$, 
and is lower than the characteristic velocity of energy injection, $A\omega$. 
This is followed by a transition region, were $\left<C\right>_t$ increases 
sharply and superlinear on $A\omega$, eventually surpassing the $A\omega$ line. 
Finally, $\left<C\right>_t$ saturates as the system enters the stable 
buoyancy-driven convective state. Quantitatively, we define the limits of the 
transition region by looking at the intersection of the initial and final linear 
behaviors with the increasing transient behaviour, the two points defining the 
width of the transition, $\delta \omega$, and their average the critical 
frequency $\omega_c$, which coincides within measurement error with the 
condition $\left<C\right>_t = A\omega$. For the narrow container, this results 
in critical frequencies and widths of transition $\omega_c = 33.4\pm0.1$, 
$\delta \omega = 0.22\pm0.01$ for EBC and PBC, and $\omega_c = 32.9\pm0.1$, 
$\delta \omega = 0.26\pm0.01$ for DBC, with the error given by the resolution of 
the simulations in $\omega$. That is, elastic boundary conditions have no 
measurable influence when compared to periodic boundaries, which suggests that 
excluded volume effects due to the presence of walls can be disregarded already 
for $l_x = 50d$. Dissipative boundaries, on the other hand, have the at first counterintuitive
effect of decreasing $\omega_c$, while increasing $\delta \omega$; 
even though overall the system presents more dissipation compared to the EBC 
case, inelastic sidewalls slightly reduce the energy needed to trigger the 
transition compared to elastic boundaries.

Boundary conditions in the wide container become irrelevant, with all cases given by $\omega_c = 33.0 \pm 0.1$ and $\delta \omega = 0.29\pm0.02$. Quantitatively, the critical points are slightly lower and the transitions wider, which we believe is due to the influence of the confinement in the narrow container. It is worthy to remark that the amount of energy needed for the creation of the convective cells is practically invariant on $l_x$ or, equivalently, the number of convective rolls $n_c \equiv 2l_x/\lambda_c$, suggesting that the interaction between rolls has no influence on their creation. Nevertheless, we notice that when EBC or DBC are used, convection cells are seen to appear first at the boundaries, and the boundary rolls are more stable when compared to the bulk of the system. This, nevertheless, happens at the same $\omega_c$ as with PBC, suggesting that solid boundaries have no relevant influence on the flux ($n\vec v$) strength, but do promote the appearance of convective cells near them.

When normalized by $A \omega$, we can recognize in $C^* \equiv \left<C\right>_t/A \omega$ a shape characteristic of a supercritical pitchfork bifurcation, as shown in the insets of Fig.~\ref{fig:ConvectionStrength}. The second branch of the ideal pitchfork supercritical bifurcation would correspond to taking the minimum in $x$, instead of the maximum, in \eqref{eq:convection_intensity}. When the bifurcation parameter $\varepsilon = (\omega-\omega_c)/\omega_c$ is used as control parameter, all three boundary condition cases coincide for all system sizes considered. This suggests that the transition presents universal behaviour, independent of the amount of dissipation. It is also a confirmation that the critical points are well defined. With the phase-space determined, next we characterize the precursors of the transition by looking first at correlations of the velocity field (\ref{sec:ConvectionTransient}), and then at density fluctuations by means of the static structure factor (\ref{sec:StaticStructure}).

\subsection{Time-dependent fluctuating convective flows}
\label{sec:ConvectionTransient}

\begin{figure}
 \begin{center}
  \includegraphics[scale=0.84]{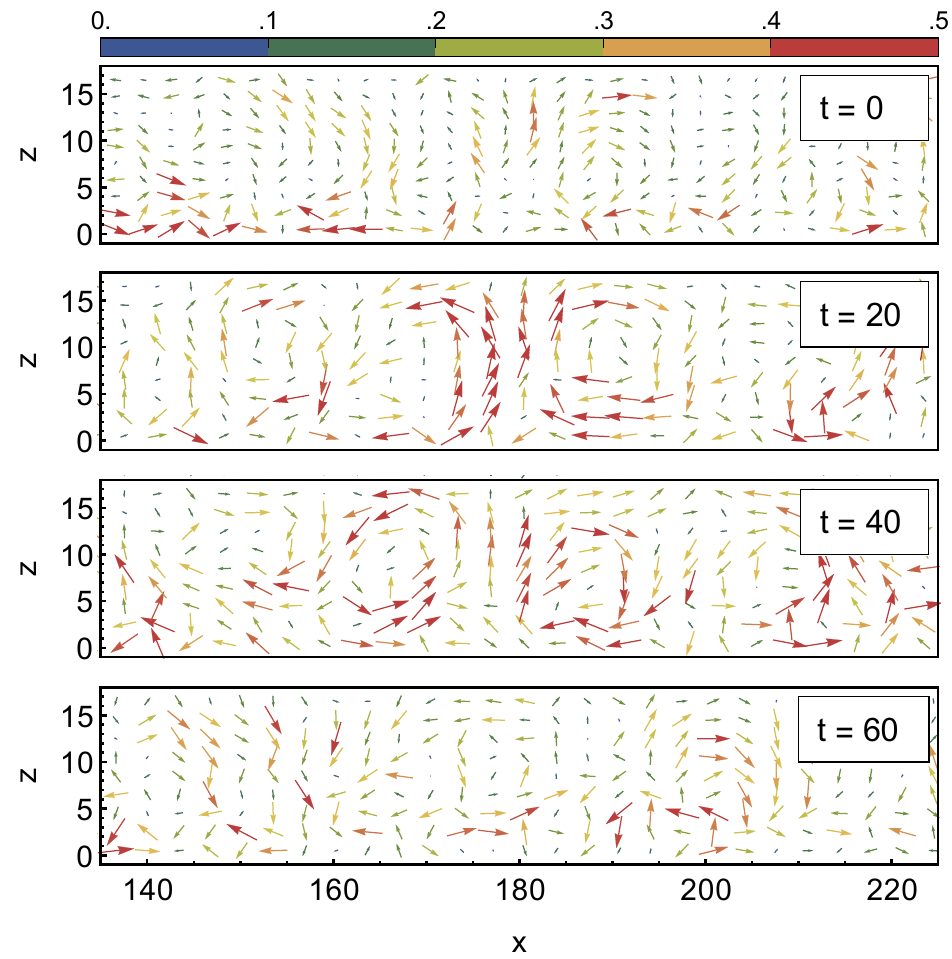}
 \end{center}
 \caption{Transient velocity fields for $\varepsilon = -0.15$, each averaged over $5$ oscillation periods, showing the emergence and decay of a fluctuating convective cell in a section of a wide container. From blue to red, the color and size of the vectors corresponds to their norm.}
 \label{fig:FluctuatingConvection}
\end{figure}

Far below the transition point in the Leidenfrost state, starting from 
$\varepsilon > -0.5$, time-dependent fluctuating convective flows can be 
observed. These are analogous to the precursor's fluxes present in the classical 
fluid Rayleigh-B\'enard convection transition \cite{bodenschatz_recent_2000}, 
which were theoretically predicted and relatively recently observed by careful 
experiments in gaseous media \cite{oh_thermal-noise_2003}. In our case, the 
convective rolls can be easily identified when observing the evolution of the 
short-time-averaged transient velocity fields, as shown in 
Fig.~\ref{fig:FluctuatingConvection}. The cells are constantly generated 
anywhere in the container, but more frequently next to walls, this of course 
when they are present, i.e. in the EBC and DBC cases. Two fundamental aspects 
differentiate such a transient state from the fully developed buoyancy-driven 
convective state above $\varepsilon = 0$. First, the circulation of particles is 
not associated with mean density or temperature inhomogeneities (it is 
time-dependent). Second, the convective velocity field is present only as an 
average, and thus is not correlated with the instantaneous velocity of the 
particles. That is, the velocities of the fluctuating convective flows are much 
smaller than the amplitude of the fluctuating velocities ($\sqrt{T}$), in 
contrast to the buoyancy-driven convection case, where they are comparable (see 
Fig.~\ref{fig:2dfields}). This has the consequence that, as there is no 
localization of the fluxes, their effect is not reflected in $\left<C\right>_t$.

\begin{figure}
 \begin{center}
  \includegraphics[scale=0.5]{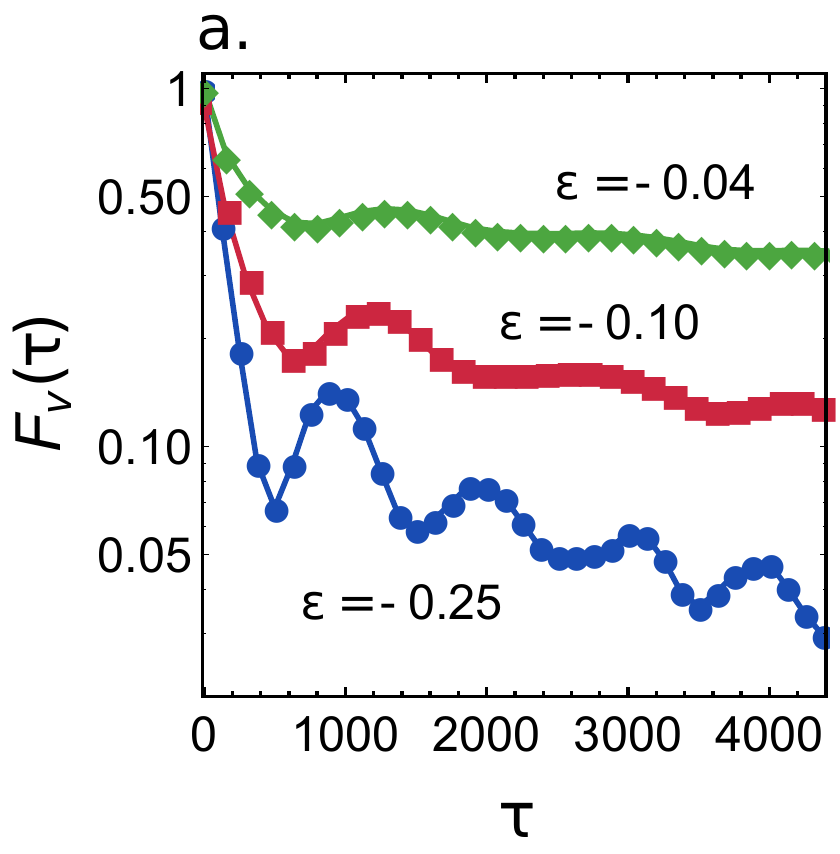}
  \includegraphics[scale=0.5]{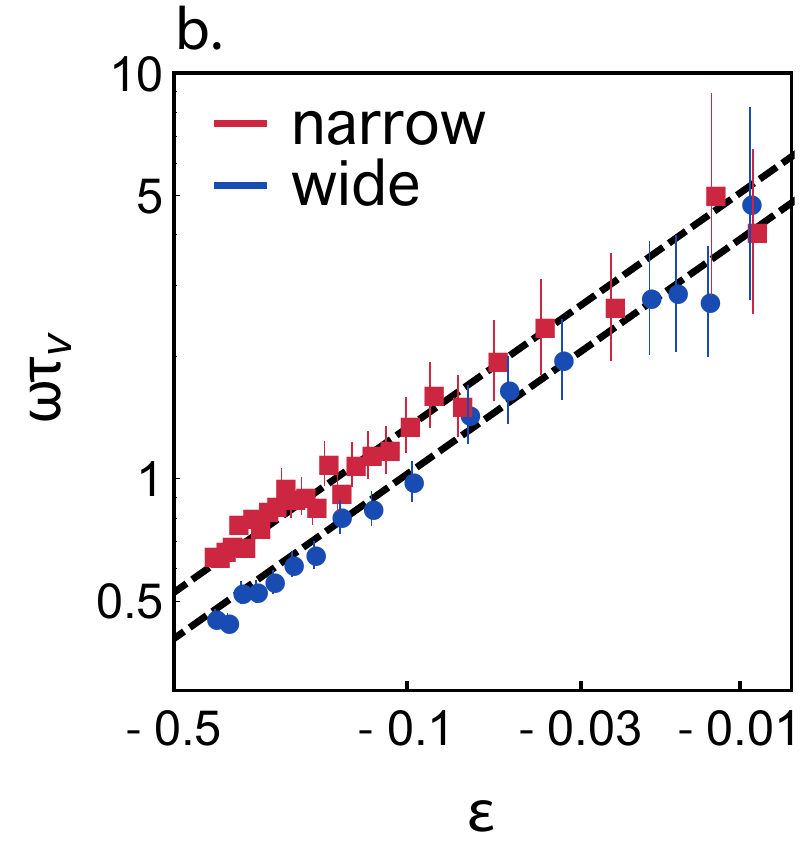}
 \end{center}
 \caption{(a) Velocity correlation functions $F_v$ for several $\omega$ and EBC in the narrow container.
	  (b) Characteristic time-scale of fluctuating convection $\tau_v$, corresponding to the exponent of the long term exponential decay of the self correlation function $F_v$, as a function of the bifurcation parameter $\varepsilon$. The dashed lines indicate best fits of the form indicated in the main text.}
 \label{fig:CorrelationTime}
\end{figure}

In order to characterize the stability of the transient convective cells, the 
self-correlation of the fluctuating velocity field is computed,
\[
 F_v(\tau) = c_F \left< \delta \vec v (\vec x, t + \tau) \cdot \delta \vec v (\vec x, t) \right>_x
\]
with $\delta v = \vec v(\vec x, t) - \left<\vec v(\vec x, t)\right>_t$, and 
$c_F$ a normalization constant such that $F_v(0) = 1$. 
Fig.~\ref{fig:CorrelationTime}a shows $F_v(\tau)$ for characteristic cases of 
$\omega$. In the following we focus only on EBC and DBC, as they considerably 
simplify the computation of self-correlation functions by impeding the 
convective rolls to drift in the $x$-direction, as they do with PBC. Visual 
inspection and preliminary analysis of the PBC case suggest that the results can 
be generalized to this case as well. All correlations present a common shape: an 
initial quick, power-law-like decay followed by a slower exponential decrease. 
The rapid decorrelation at short time-scales confirms that the particles' 
instant velocities are mostly fluctuating, and do not present a high time 
correlation. On the other hand, for longer times the correlation is 
comparatively lower, but still considerable, and decays slower. This is a signal 
of long-term average preferred fluxes. As expected by the critical slowing down 
of fluctuations near the transition, the overall correlation of this region 
increases as the critical point is approached, as can also be seen in 
Fig.~\ref{fig:CorrelationTime}a. The characteristic time of decorrelation 
$\tau_v$ is obtained by considering $F_v \sim \exp(-\tau/\tau_v)$. 
Fig.~\ref{fig:CorrelationTime}b shows $\omega \tau_v$ as a function of 
$\varepsilon$, from where we find a powerlaw $\tau_v \sim 
\varepsilon^{-\xi}/\omega$ with exponent $\xi \sim 0.59\pm0.02$. Closer to the 
critical point the measurement error becomes significant. The data is presented 
for the whole range in $\omega$ where the Leidenfrost state is present, which is 
one and a half decades in $\varepsilon$.

Wide systems present the same overall features as the narrow container; 
$\tau_v$ can be determined with a higher precision --as noise is reduced with a 
higher number of particles--, and presents the same, within error, critical exponent $\xi$ as 
in the narrow case, $\xi \sim 0.60\pm0.02$. That is, transient convective flows are independent of the 
size of the container.

Also visible in $F_v(\tau)$ are wide peaks at regular intervals, signals of a 
quasi-periodic time-scale of correlation. By observing the evolution of the 
center of mass, and computing its fast-Fourier transform, it was verified that 
this periodic correlation corresponds to the recently reported low-frequency 
oscillations, present in density inverted agitated 
systems~\cite{rivas_low-frequency_2013, windows-yule_low-frequency_2014}. The 
quasi-periodic movement is coupled with a breathing behaviour of the dense bed 
of grains, which increases and decreases its granular temperature. Here we do 
not analyze this further; for a detailed study of the phenomena we refer the 
reader to~\cite{rivas_low-frequency_2013}, and a further experimental study 
in~\cite{windows-yule_low-frequency_2014}.

\subsection{Static structure function}
\label{sec:StaticStructure}

As energy input increases, for $\varepsilon > -0.1$, density fluctuations arise, clearly recognizable as modulations in the surface of the bed of grains. To analyze their behaviour we compute the static structure function,
\begin{align}
	\staticStructure(k) = \frac{1}{N} \left< \left| \hat n(k,t) - \left< \hat n(k,t) \right>_t \right|^2 \right>_t,
\end{align}
with $\hat n$ the Fourier components of the depth-averaged number density field in the $x$-direction,
\begin{align}
	\hat n(k, t) = \sum_j^{l_x/\delta_x} n(x_j, t) e^{i 2 \pi k \, n(x_j, t)}.
\end{align}
Notice that we define $k = 1/\lambda$, for a more straightforward comparison 
between wave number $k$ and wavelength $\lambda$. The position $x_j$ is given by 
regular intervals, $x_j = \tfrac{1}{2} \delta x + j \delta x$, with $\delta x = 
0.1$ the coarse graining length. Notice that instead of considering the 
particles' position in the definition of $\hat n$ we use the averaged density 
profiles, as it significantly increases the speed of computation. This 
approximation holds only for low wave-numbers, that is, $1/k \gg \delta x$, 
which is the region we are interested in. Test cases were done with the usual 
definition with particle positions, and no significant differences were 
observed. 

\begin{figure}
 \begin{center}
  \includegraphics[scale=0.5]{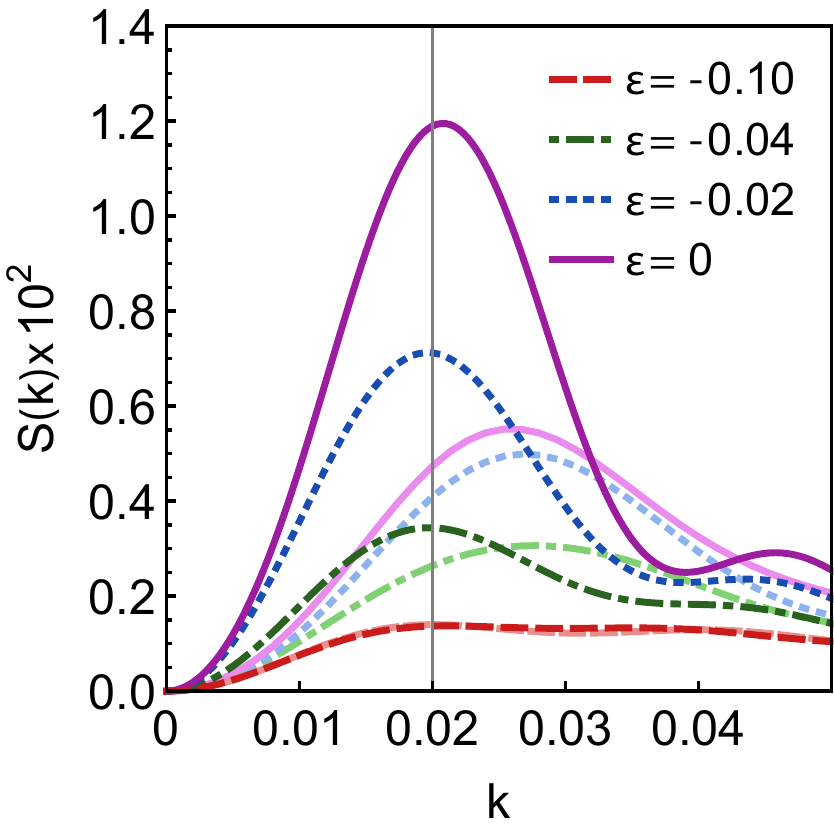}
  \includegraphics[scale=0.5]{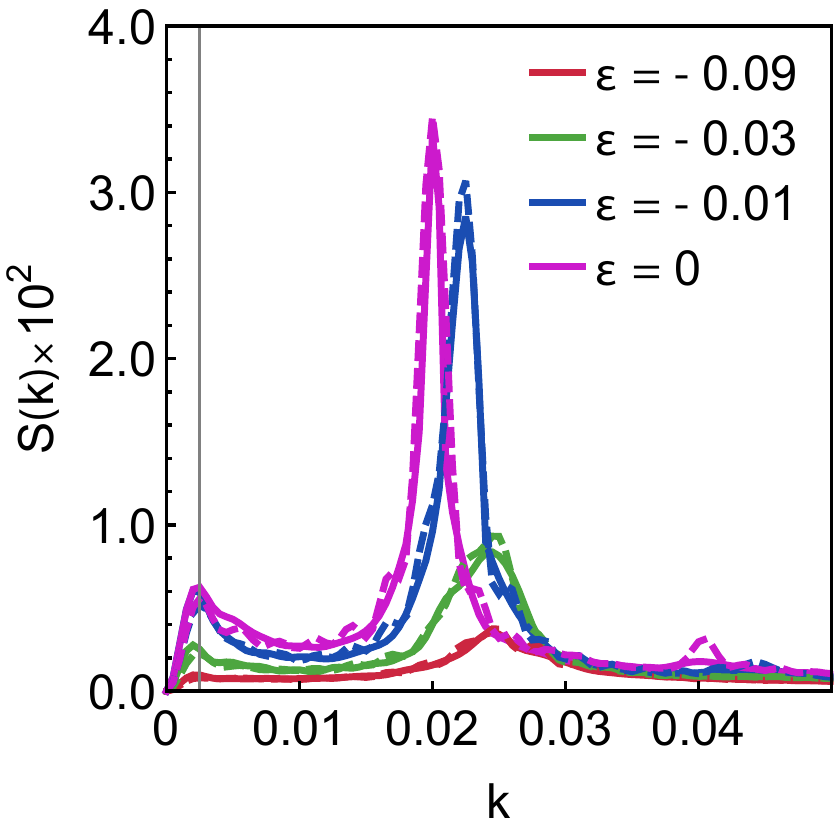}
 \end{center}
 \caption{Structure factor, $S(k)$, for narrow (left) and wide (right) containers for the bifurcation parameters specified. Dashed lines correspond to PBC, while solid lines have EBC. The vertical solid line indicates the $1/l_x$ point.}
 \label{fig:StructureFactor}
\end{figure}

Transient modulations of the bed are captured in $S(k)$ by the appearance and 
steady increase of a narrow peak at $k \approx 0.02$, as shown in 
Fig.~\ref{fig:StructureFactor} for both the narrow and wide containers. We 
define the critical mode $k_c$ by the position of this maximum, that is $S_m 
\equiv \max(S(k)) \equiv S(k_c)$. Thus, the associated wavelength at the 
transition point, $\lambda_c^* \approx 50$, corresponds to the size of the 
smallest stable convection roll, seen to be independent of $l_x$ for $l_x > 
\lambda_c^*$. Notice that this corresponds to $n_c = 2$ for the narrow container, and $n_c = 8$ for the wide container. 
What other factors may affect $\lambda_c^*$ is not studied further 
here, although we notice that previously realized stability analysis of the 
granular hydrodynamic equations have found an expression for $\lambda_c^*$ as a 
function of the constitutive relations, which are in themselves dependent on the 
particle properties \cite{eshuis_buoyancy_2013}.

Notice from Fig.~\ref{fig:StructureFactor} that for $\varepsilon = -0.1$ the 
correlation of the transient convective flows was significant, but $S(k)$ has no 
relevant maximum. This confirms that fluctuating convective flows take place in 
a stable homogeneous Leidenfrost state, and are not accompanied by any relevant 
excitation of the critical mode in the density (and temperature) field.

\begin{figure}
 \begin{center}
  \includegraphics[scale=0.5]{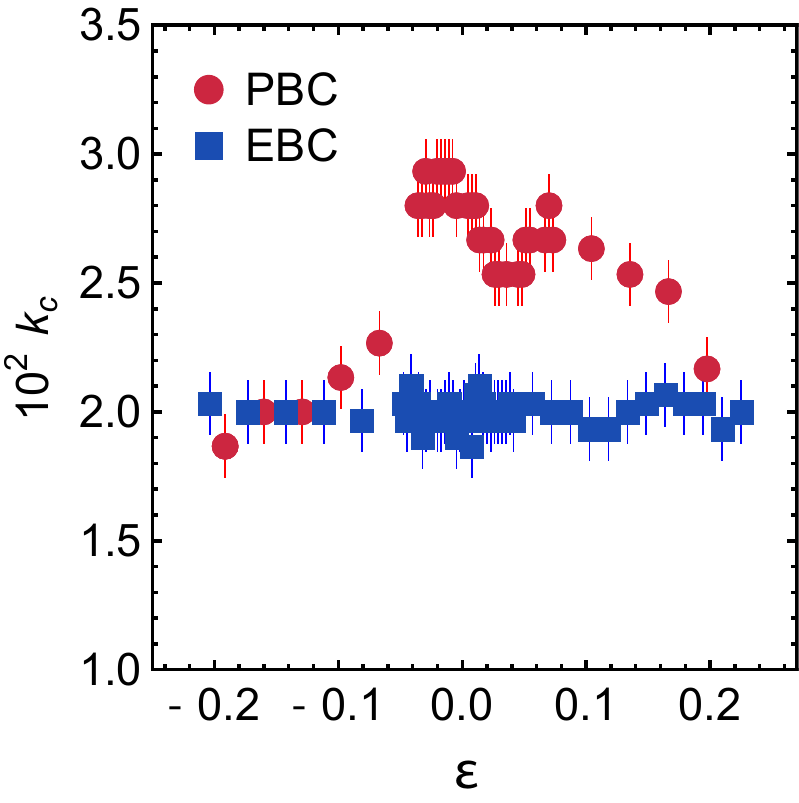}
  \includegraphics[scale=0.5]{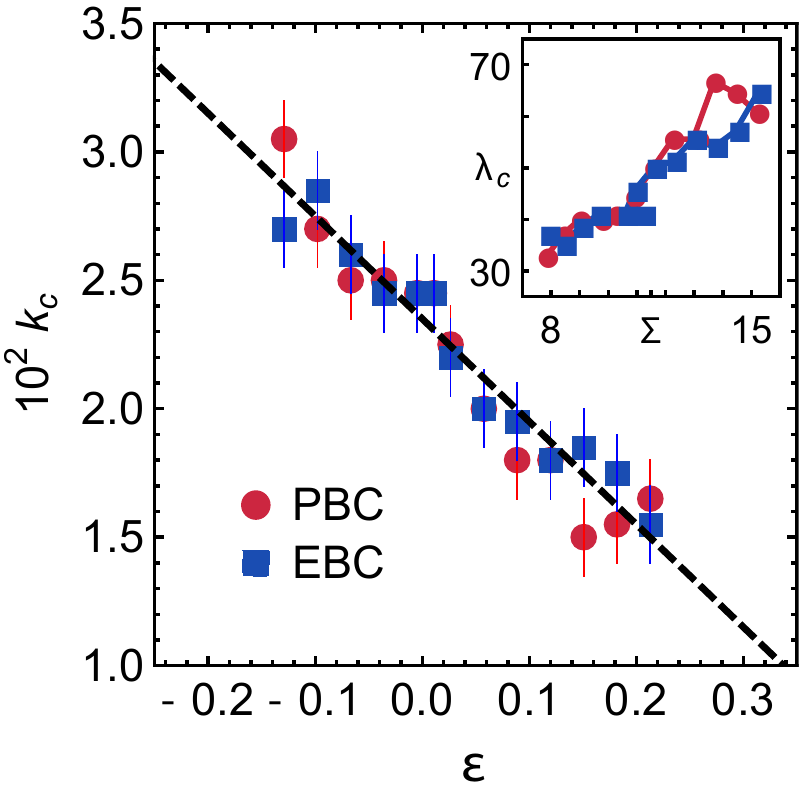}
 \end{center}
 \caption{The most unstable mode $k_c$, defined by the maximum of the structure factor $\max(S(k)) \equiv S(k_c)$, as a function of the bifurcation parameter, for narrow (left) and wide (right) containers and the boundary conditions specified. The inset shows the convective length-scale $\lambda_c$ as a function of the shaking strength, as defined in the main text.}
 \label{fig:komega}
\end{figure}

Previous simulational and experimental works have stated that $\lambda_c$ 
scales linearly with the shaking strength $\Sigma \equiv \tilde A^2 \tilde \omega^2/gd = A^2 \omega^2$~\cite{luding_simulations_1996, eshuis_onset_2010}. 
The inset of Fig.~\ref{fig:komega} shows $\lambda_c(\Sigma)$ for the wide 
container and confirms that this is indeed the case. 
We cannot distinguish any effect of confinement, which could be identified as plateaus in the increase of $\lambda_c$; this is to be expected in the $\lambda_c \ll l_x$ limit, which is the case for the wide container.
In contrast, in the narrow 
container solid walls fix $k_c$, while with PBC the behaviour is not clear, 
roughly increasing before the transition point and then decreasing in a 
non-monotonic way; the uncertainty in the measurements does not allow a more 
accurate conclusion in this case. The marked difference between both boundary 
condition cases suggests that, even though EBC and PBC had equal critical 
points, as measured by $\left<C\right>_t$, they do have an influence on the 
modes that are being perturbed. In most of the studied range $k^*$ is 
consistently higher with PBC, showing that solid boundaries can have the 
originally unexpected effect of increasing the critical convection roll size. 
This is due to excluded volume effects near the wall, which decrease the density 
and thus have the effect of exciting a lower mode, in our case for $\lambda 
\approx 25$. On the wide container wall effects become negligible, and thus 
$k^*$ coincides for both types of boundary conditions.

By taking into account that $k_c$ in the wide container is not constant, we interpret the LBC transition for $l_x > \lambda_c^*$ as a series of transitions between energetically similar states. Inherent fluctuations are strong enough to allow the constant switching between contiguous $k_c$. In terms of the relevant scales, this is a conflict between $\lambda_c$, which depends on our control parameter $\omega$, and $l_x$, which is fixed. As $\lambda_c^*$ is independent of the container size, the critical behaviour for $|\varepsilon| \sim 0$ is still expected to be universal.

\begin{figure}
 \begin{center}
  \includegraphics[scale=0.5]{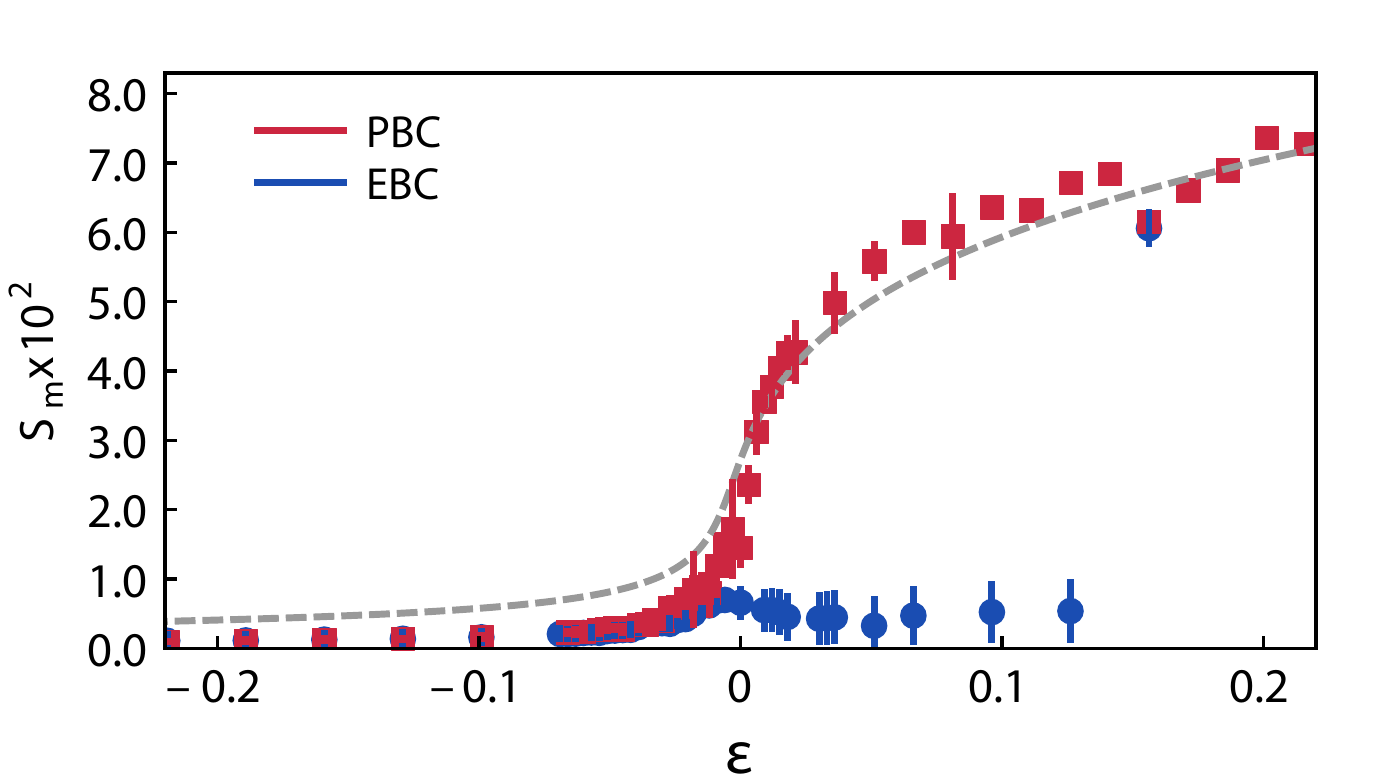}
  \includegraphics[scale=0.5]{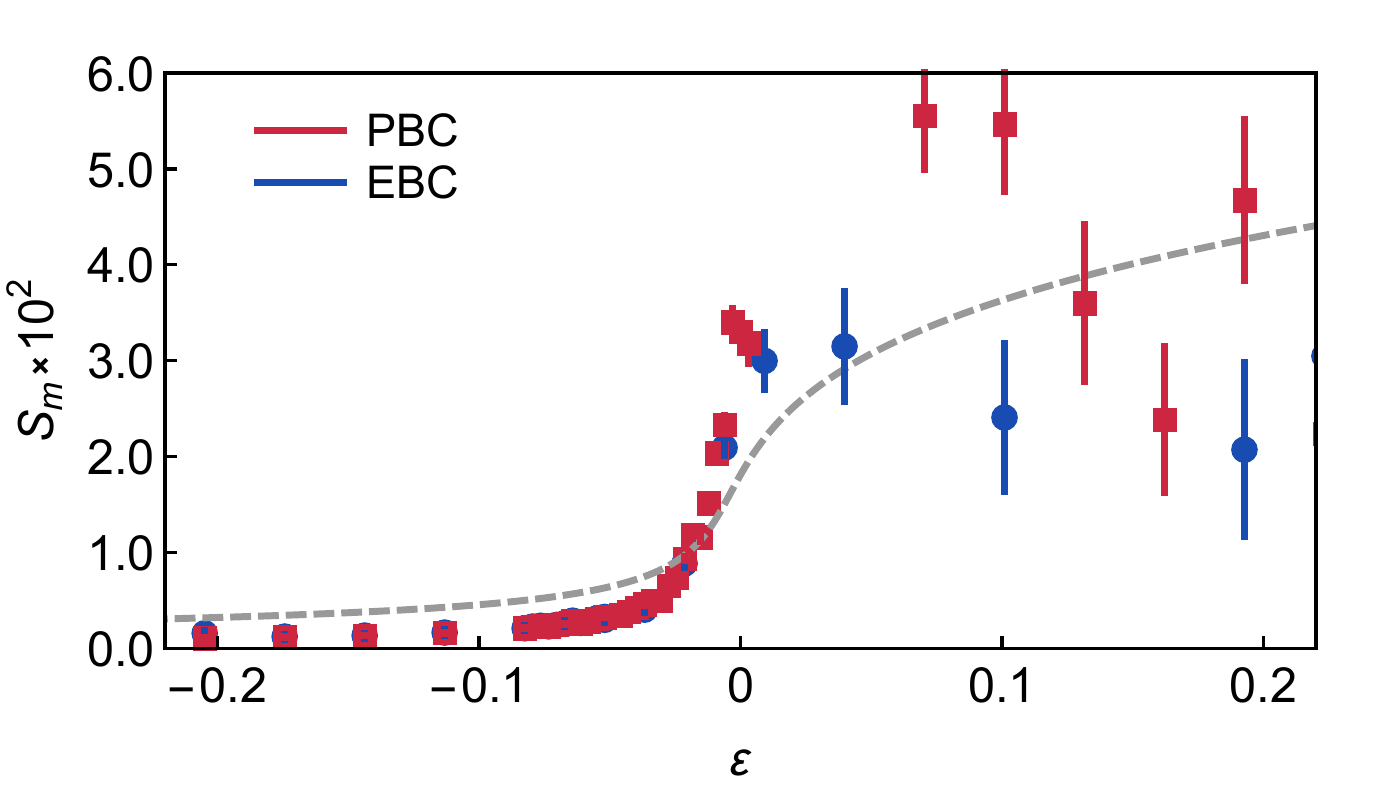}
 \end{center}
 \caption{Structure factor maximum, $S_m$, as a function of the bifurcation parameter $\varepsilon$, for EBC (blue) and PBC (red) in narrow (top) and wide (bottom) containers. As reference, the best fit for the amplitude of the critical mode is included (dashed gray, see main text).}
 \label{fig:StructureFactorMax}
\end{figure}

Indeed, $S_m(\varepsilon)$ shows critical-like behaviour for $\varepsilon < 0$, as shown in Fig.~\ref{fig:StructureFactorMax}. In the narrow container, both types of boundary conditions show the same qualitative growth for $\varepsilon < 0$, although with consistently lower amplitudes in the EBC case, as previously discussed. For $\varepsilon > 0$ the PBC case shows a growth reminiscent of the critical amplitude of a supercritical bifurcation. In this case, $S_m$ is directly related to the amplitude of the critical mode, as the lack of a fixed reference frame makes $\left<n(x,t)\right>_t$ homogeneous even in the buoyancy-driven convective state. On the contrary, the EBC case immediately decays for $\varepsilon > 0$. In the wide container both cases coincide within error for $\varepsilon < 0$, showing that the discrepancy between both cases in the small container is indeed a size-effect. $S_m$ again looses significance for $\varepsilon > 0$, and the behaviour is erratic due to metastability of the transient region in the wide systems.

\begin{figure}
 \begin{center}
 \includegraphics[width=0.48\textwidth]{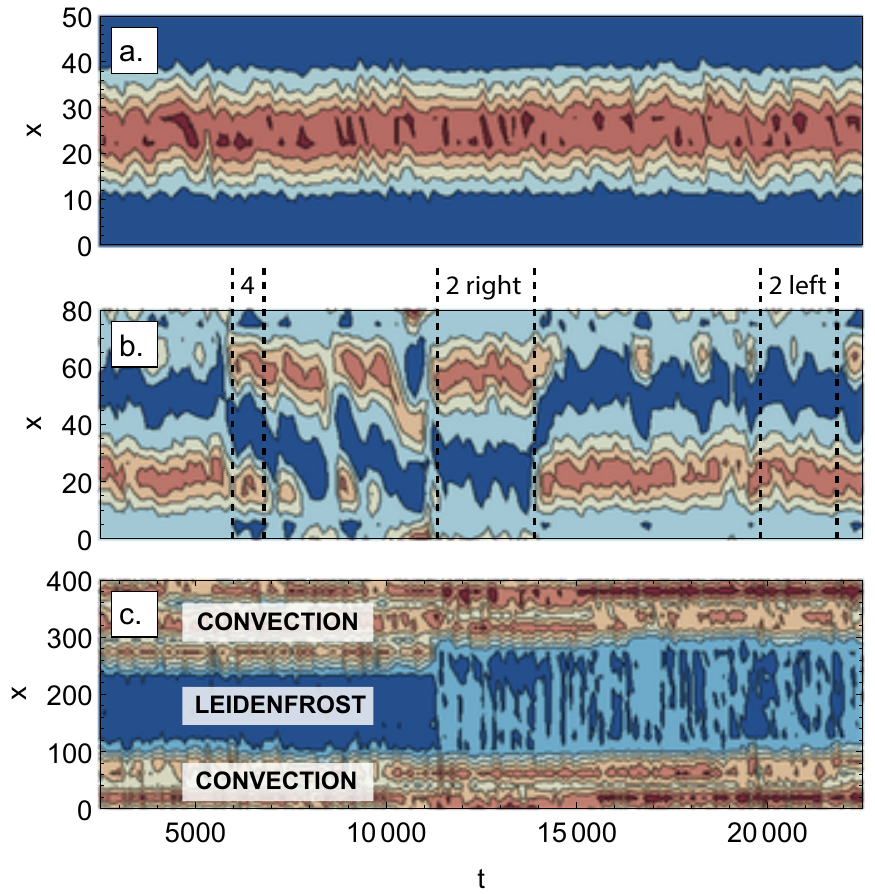}
 \end{center}
 \caption{Spatio-temporal contours plot of the number of particles field, $n(x,t)$, for (a) $l_x = 50$, (b) $l_x = 80$ and (c) $l_x = 400$, with $\omega = 35$ ($\varepsilon \sim 0.06$) and EBC. These correspond to $l_x/\lambda_c^* \sim 1$, $l_x/\lambda_c^* \sim 1.6$ and $l_x/\lambda_c^* \sim 8$, respectively. High density regions are shown in red. Over the middle figure, the number of convection rolls is indicated for exemplary regions.}
 \label{fig:SpatioTemporal}
\end{figure}

\subsection{Dynamics of transient states}

The buoyancy-driven convective state presents complex time evolutions. These 
are heavily dependent on the $l_x/\lambda_c$ ratio, based on the constraint that 
the number of convective rolls has to be an integer number. Half-integer values 
are possible only with solid wall boundary conditions. This implies that 
non-integer values of $l_x/\lambda_c$ lead to metastable states, as the number 
of convective cells $n_c$ presents intermittent behaviour between the two 
closest values of $k_c$, as also between convection rolls at different sides of 
the container, if walls are present. As an example, 
Fig.~\ref{fig:SpatioTemporal}b shows the temporal evolution of $n(x)$ for a 
system with $l_x = 80$, that is, $l_x/\lambda_c^* \approx 1.6$. For a value of 
$\omega$ just after the transition point, the convective cell constantly 
switches between metastable states; it is possible to identify two-rolls and 
one-roll configurations at either side of the system, alternating with no clear 
periodicity. We believe this to be an important factor to take into account on 
any study of the dynamics of the granular convective state: the size of the 
container has no influence on the critical point of the transition, but plays a 
determining role in the dynamics. In our case, $l_x$ for the narrow and wide 
containers was chosen a posteriori to diminish the effects of metastability, 
considerably facilitating the study of precursors.

As $l_x/\lambda_c$ is increased further, a new state becomes possible at the 
transition region in which convective cells coexist with regions effectively in 
the Leidenfrost state. Fig.~\ref{fig:SpatioTemporal}c shows a period of 
coexistence, as two pairs of convective cells emerge in a confined region of the 
system while the rest remains in the Leidenfrost state. Notice how the 
Leidenfrost region is roughly $200d$ wide, far larger than $\lambda_c^*$. We 
interpret this phenomena as the emergence of a localized state in a non-linear 
system, a subject of increased scientific interest~\cite{dawes_emergence_2010}.

\subsection{Critical mode amplitude}
\label{sec:Amplitude}

It has been shown that both the correlation of the fluctuating velocity field 
and the critical mode of the density fluctuations present critical-like 
behaviour near the transition. We now look at the overall transition behaviour 
in the context of bifurcation theory, by following the amplitude in the emergent 
pattern of the critical mode, $A_c$. The emergent pattern is more evident and 
measured from the vertical velocity field $v_z(x,t)$. $A_c$ is the amplitude of 
the mode $k_c$ in $v_z(x,t)/\omega$, with $k_c$ determined by the structure 
factor maximum. The final value of $\left<A_c\right>_t$ is obtained by averaging 
over the whole simulation time.

In the seminal work of Swift and Hohenberg, hydrodynamic fluctuations were 
studied for a molecular fluid near the thermal convection instability 
\cite{swift_hydrodynamic_1977}, and a simple model for the Rayleigh-B\'enard 
instability was derived. In the following we apply the Swift-Hohenberg model to 
the LBC transition, inspired by the evident similarities of both phenomena; in 
terms of bifurcation theory, both transitions correspond to spatial-mode 
selecting bifurcations. Nevertheless, we expect the discrete nature of our 
granular system to have a considerable effect close to the transition, 
manifested as fluctuations arising from the finite number of particles. Thus, we 
consider that the universal behaviour of the fluctuating vertical velocity 
$w(z,t) = v_z(x,t) - \left<v_z(x,t)\right>_t$ close to the transition is given 
by the Swift-Hohenberg model for pattern formation with a stochastic 
term~\cite{garcia-ojalvo_noise_1999},
\begin{equation}
	\partial_t w = \varepsilon' w - w^3 - (\partial_{xx} + k_c^2)^2 w + \sqrt{\eta'} \zeta(x,t),
	\label{eq:normalform}
\end{equation}
with the bifurcation parameter given by $\varepsilon' - k_c^4$. In our system 
$\varepsilon' - k_c^4 \approx \varepsilon$ (as $k_c \ll 1$), and thus in what 
follows we take $\varepsilon' = \varepsilon$. Fluctuations are modeled by the 
last term, where $\zeta$ is a Gaussian white noise, that is 
$\left<\zeta(x,t)\zeta(x',t')\right> = \delta(x-x')\delta(t-t')$; and $\eta'$ is 
the parameter of noise intensity~\cite{agez_universal_2008}. In our system the 
zero correlation of $\zeta$ is justified by assuming the gaseous phase close to 
the moving plate to be the main source of fluctuations, and to behave strictly 
as a hard sphere gas. The lack of temporal or spatial correlations of the 
particles follows from the the low packing fractions ($\phi < 0.2$), the frequent collisions with the bottom plate compared to the mean free flight time, and the randomization of velocities due to collision with the dense region.

\begin{figure}
 \begin{center}
  \includegraphics[scale=0.5]{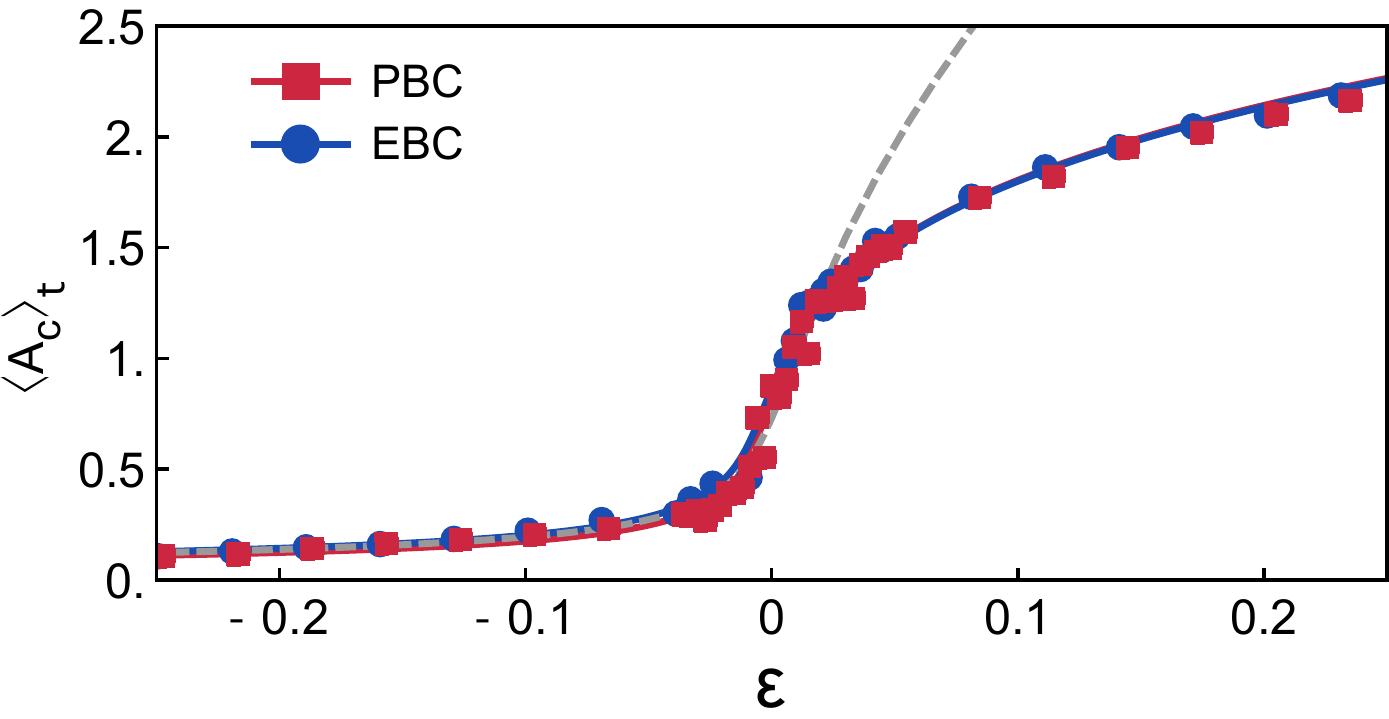}
  \includegraphics[scale=0.5]{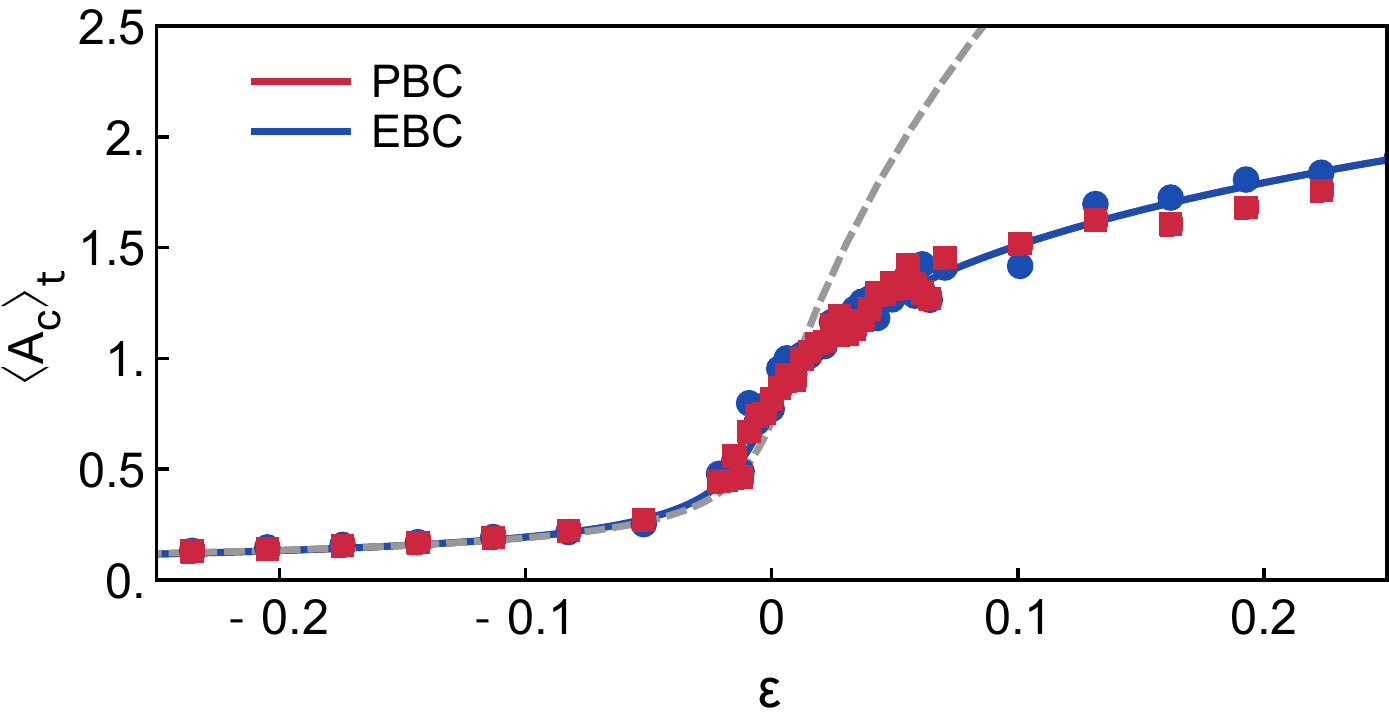}
 \end{center}
 \caption{Amplitude of the critical pattern of the vertical velocity field $v_z(x,t)$, $A_c$, as a function of the bifurcation parameter $\varepsilon$, for EBC (blue) and PBC (red) in narrow (top) and wide (bottom) containers. The gray dashed lines correspond to fits given by the Swift-Hohenberg model with a stochastic term (see main text), with noise level $\eta = 0.0001$. The colored dashed lines correspond to fits based on a quintic supercritical bifurcation, for noise intensity $\sigma = 0.0008$ in the small container systems, and $\sigma = 0.001$ for the wide cases.}
 \label{fig:Amplitude}
\end{figure}

It is known that in \eqref{eq:normalform} the base state $w(x,t) = 0$ is stable for $\varepsilon' < 0$, and presents a supercritical spatial instability for $\varepsilon' = 0$, which leads to the appearance of a pattern, in our case corresponding to convective cells, for $\varepsilon' > 0$. Following~\cite{agez_universal_2008,agez_bifurcations_2013}, and confirmed by our measured velocity profiles, solutions for the critical mode $k_c$ can be assumed to be of the form
\begin{equation}
	w = \frac{a(\tau)}{\sqrt{3}} e^{ik_cx} + \frac{\bar a(\tau)}{\sqrt{3}} e^{-ik_cx} + U(a,\bar a, x)
	\label{eq:solution}
\end{equation}
with $a$ the amplitude of the pattern with mode $k_c$, dependent on the slow time $\tau \equiv \varepsilon t$, and $U$ a general function containing higher order terms in $a$. Substituting \eqref{eq:solution} into \eqref{eq:normalform} one reaches the amplitude equation corresponding to a stochastic cubic supercritical spatial bifurcation:
\begin{equation}
 \partial_\tau a = \varepsilon a - | a |^2 a + \sqrt{\eta} \zeta(\tau)
\end{equation}
with $\eta \equiv 3 \eta'$. A solution for the probability function of $a$, $P_s(|a|,\varepsilon,\eta)$, can be found from \eqref{eq:normalform} and \eqref{eq:solution}, as shown in~\cite{agez_universal_2008,agez_bifurcations_2013}. From the shape of $P_s$ the expectation value can be obtained~\cite{agez_universal_2008}, given by
\begin{equation}
  \left|a_{\text{max}}\right| = \sqrt{\frac{\varepsilon + \sqrt{\varepsilon^2 + 2\eta}}{2}}.
  \label{eq:a_max}
\end{equation}
In our case $\left|a_{\text{max}}\right| = \left<A_c\right>_t$. Our measurements are consistent with this form for $|\varepsilon| \ll 1$, as shown in Fig.~\ref{fig:Amplitude} for narrow and wide systems with PBC and EBC. Nevertheless, \eqref{eq:a_max} does not capture the shape of $\left<A_c\right>_t(\varepsilon)$ for higher values of~$\varepsilon$, deviating considerably already for $\varepsilon \sim 0.05$.

A higher level of agreement can be obtained by considering an stochastic quintic supercritical bifurcation~\cite{agez_bifurcations_2013}:
\begin{equation}
  \partial_\tau a = \varepsilon a - | a |^4 a + \sqrt{\eta\sqrt{h}} \zeta(\tau),
  \label{eq:quintic}
\end{equation}
with $h$ quantifying the strength of the quintic non-linear term~\cite{agez_bifurcations_2013}. This type of bifurcation may be more relevant for our system, as it has been previously associated with parametrically driven spatially extended systems, as Faraday patterns~\cite{coullet_dispersion-induced_1994} and vertically vibrated series of coupled pendula \cite{clerc_localized_2008}. The former can be considered closer to our present system than the Rayleigh-B\'enard scenario, taking into account that the bed of grains is also a vertically vibrated medium with a free surface. The latter case, on the other hand, could be related to the already mentioned low-frequency oscillations~\cite{rivas_low-frequency_2013}. Previously it was shown that density inverted granular states in a quasi-one-dimensional container ($\tilde l_x \sim \tilde l_y \sim d$) behave approximately as harmonic oscillators. It can thus be inferred that for wider containers, as the ones considered in this study, the dynamics are analogous to a series of coupled oscillators, with a yet unspecified coupling mechanism by shear
or other interactions.

Following a similar method as the previous analysis, an expression for the expected value of the amplitude of the unstable mode can be obtained (for details of the derivation we again refer the reader to~\cite{agez_bifurcations_2013}),
\begin{equation}
  \left|a_{\text{max}}\right| = \sigma^{1/6} \sqrt{\beta/\Omega + \Omega/3}
  \label{eq:a_max5}
\end{equation}
with $\Omega \equiv (3/4)^{1/3}(9+\sqrt{3(27-16\beta^3)})^{1/3}$ and $\beta \equiv \varepsilon/\sigma^{2/3}$, with $\sigma \equiv \eta \sqrt{h}$. The shape of \eqref{eq:a_max5} is also shown in Fig.~\ref{fig:Amplitude}, now in good agreement for higher $\varepsilon$ in all cases.

All systems present the same overall shape of $A_c(\varepsilon)$, with the most significant difference being lower amplitudes for $\varepsilon > 0$ in the wide containers. More importantly, there is no significant difference in the noise intensity for all cases, except for the narrow container with EBC, where the noise term is lower, $\sigma = 6\times10^{-4}\pm10^{-5}$. In the narrow container with PBC $\sigma = 9\times10^{-4}\pm2\times10^{-4}$, and $\sigma = 10^{-3}\pm2\times10^{-4}$ for the wide container with any boundary condition. The independence of the noise intensity on $N$ suggests that the relevant quantity for the critical dynamics is the amount of particles per critical length-scale, $\lambda_c^*$, which in our cases remains constant.

\section{Conclusions}

We have studied the granular Leidenfrost to buoyancy-driven convection 
transition, characterized the precursors and proposed a new interpretation 
of its universal dynamical behaviour. The overall picture is of a continuously 
fluidized bed of grains which goes from homogeneous Leidenfrost configurations 
to increasingly velocity correlated convective states, until flows are strong 
enough to sustain the density inhomogeneous buoyancy-driven convective state. 
From a bifurcation theory perspective, the convection transition can be 
understood as a pattern formation phase-transition, with the emergence of 
convective cells with a critical length-scale independent of the domain size, 
which is consistent with previously realized hydrodynamic stability analysis of 
the Leidenfrost state \cite{eshuis_buoyancy_2013}.

The time-dependent fluctuating convection state can be characterized by the 
correlation time of the fluctuating velocity field, which shows critical-like 
behaviour with an exponent of approximately $0.51$. From the self-correlation it 
is also possible to observe the influence of low-frequency oscillations 
\cite{rivas_low-frequency_2013} on the fluctuating velocity field.

The static structure factor shows the emergence and growth of the pattern dominant length-scale. The amplitude of the critical mode is also seen to show critical behaviour, consistent with a supercritical bifurcation. By following the most unstable mode throughout the transition in wide systems it was possible to confirm that the size of the convective cells is indeed proportional to the frequency of energy injection.

In the transient state of wider systems the Leidenfrost and buoyancy-driven convective states can coexist. The convective state in this region is constantly evolving, presenting metastability between states with different number of rolls. As energy increases the stability of the convective cells increases, although their number is determined by the amount of cells that can be fitted in the container. Further increasing the energy leads to a comparatively slower process of merging of convective cells. The rich  dynamics of merging and splitting of convective cells in coexistence with the Leidenfrost state in the wide systems calls for further research.

Elastic walls and periodic boundaries present the same critical points, disregarding any significant confinement effects for containers much bigger than the critical convective wavelength (i.e. larger than $50$ particle diameters in the cases studied). Slightly dissipative side-walls, on the other hand, have the effect of decreasing the amount of energy needed to trigger the transition, showing that the excitation of the unstable mode at the boundaries has a more significant effect than the added dissipation. In systems 400 particle diameters wide, in all cases studied, the boundary conditions
did not have any visible influence.

The amplitude of the critical mode of convection is seen to be coherent with a quintic supercritical amplitude equation. The agreement is much better than with a cubic supercritical bifurcation, associated with the Swift-Hohenberg equation. This suggests a new interpretation of the transition, closer to spatially extended parametrically driven systems than to Rayleigh-B\'enard convection. We hypothesize that the source of the parametric driving is not the vibration of the container (which has too low amplitude and high frequency to couple with the bed dynamics), but the low-frequency oscillations present in a density inverted bed of grains, i.e. the granular Leidenfrost state. In general, we remark that the universal behaviour of the density field can only be captured by considering a noise term in the corresponding amplitude equation which quantifies the discrete, finite-number effects. The noise intensity is seen to be independent on the system size, except in the confined small container. This suggests that the transition in wider systems is a local phenomenon, with the size of the critical convective cell as relevant length-scale.

As outlook, a derivation of the quintic supercritical amplitude equation from a series of coupled oscillators with the form derived in \cite{rivas_low-frequency_2013} would be a way of confirming the proposed amplitude equation.

\subsection*{Acknowledgements}

This work was financially supported by the NWO-STW VICI grant number 10828.

\bibliography{Library}{}

\begin{thebibliography}{10}

\bibitem{douady_subharmonic_1989}
S.~Douady, S.~Fauve, and C.~Laroche, ``Subharmonic instabilities and defects in
  a granular layer under vertical vibrations,'' {\em {EPL} (Europhysics
  Letters)}, vol.~8, p.~621, Apr. 1989.

\bibitem{olafsen_clustering_1998}
J.~S. Olafsen and J.~S. Urbach, ``Clustering, order, and collapse in a driven
  granular monolayer,'' {\em Physical Review Letters}, vol.~81, pp.~4369--4372,
  Nov. 1998.

\bibitem{clerc_liquidsolid-like_2008}
M.~G. Clerc, P.~Cordero, J.~Dunstan, K.~Huff, N.~Mujica, D.~Risso, and
  G.~Varas, ``Liquid–solid-like transition in quasi-one-dimensional driven
  granular media,'' {\em Nature Physics}, vol.~4, pp.~249--254, Mar. 2008.

\bibitem{ristow_pattern_2000}
G.~H. Ristow, {\em Pattern Formation in Granular Materials}.
\newblock 2000.

\bibitem{ottino_mixing_2000}
J.~M. Ottino and D.~V. Khakhar, ``Mixing and segregation of granular
  materials,'' {\em Annual Review of Fluid Mechanics}, vol.~32, no.~1,
  pp.~55--91, 2000.

\bibitem{aranson_patterns_2006}
I.~S. Aranson and L.~S. Tsimring, ``Patterns and collective behavior in
  granular media: Theoretical concepts,'' {\em Reviews of Modern Physics},
  vol.~78, pp.~641--692, June 2006.

\bibitem{muzzio_powder_2002}
F.~J. Muzzio, T.~Shinbrot, and B.~J. Glasser, ``Powder technology in the
  pharmaceutical industry: the need to catch up fast,'' {\em Powder
  Technology}, vol.~124, pp.~1--7, Apr. 2002.

\bibitem{coussot_rheometry_2005}
P.~Coussot, {\em Rheometry of Pastes, Suspensions, and Granular Materials:
  Applications in Industry and Environment}.
\newblock John Wiley \& Sons, June 2005.

\bibitem{antony_granular_2004}
S.~J. Antony, W.~Hoyle, and Y.~Ding, {\em Granular Materials: Fundamentals and
  Applications}.
\newblock Royal Society of Chemistry, Jan. 2004.

\bibitem{ramirez_thermal_2000}
R.~Ramírez, D.~Risso, and P.~Cordero, ``Thermal convection in fluidized
  granular systems,'' {\em Physical Review Letters}, vol.~85, pp.~1230--1233,
  Aug. 2000.

\bibitem{melo_transition_1994}
F.~Melo, P.~Umbanhowar, and H.~L. Swinney, ``Transition to parametric wave
  patterns in a vertically oscillated granular layer,'' {\em Physical Review
  Letters}, vol.~72, pp.~172--175, Jan. 1994.

\bibitem{thoroddsen_granular_2001}
S.~T. Thoroddsen and A.~Q. Shen, ``Granular jets,'' {\em Physics of Fluids
  (1994-present)}, vol.~13, pp.~4--6, Jan. 2001.

\bibitem{royer_high-speed_2009}
J.~R. Royer, D.~J. Evans, L.~Oyarte, Q.~Guo, E.~Kapit, M.~E. Möbius, S.~R.
  Waitukaitis, and H.~M. Jaeger, ``High-speed tracking of rupture and
  clustering in freely falling granular streams,'' {\em Nature}, vol.~459,
  pp.~1110--1113, June 2009.

\bibitem{olafsen_two-dimensional_2005}
J.~S. Olafsen and J.~S. Urbach, ``Two-dimensional melting far from equilibrium
  in a granular monolayer,'' {\em Physical Review Letters}, vol.~95, p.~098002,
  Aug. 2005.

\bibitem{goldman_noise_2004}
D.~I. Goldman, J.~B. Swift, and H.~L. Swinney, ``Noise, coherent fluctuations,
  and the onset of order in an oscillated granular fluid,'' {\em Physical
  Review Letters}, vol.~92, p.~174302, Apr. 2004.

\bibitem{ortega_subharmonic_2010}
I.~Ortega, M.~G. Clerc, C.~Falcón, and N.~Mujica, ``Subharmonic wave
  transition in a quasi-one-dimensional noisy fluidized shallow granular bed,''
  {\em Physical Review E}, vol.~81, p.~046208, Apr. 2010.

\bibitem{agez_bifurcations_2013}
G.~Agez, M.~G. Clerc, E.~Louvergneaux, and R.~G. Rojas, ``Bifurcations of
  emerging patterns in the presence of additive noise,'' {\em Physical Review
  E}, vol.~87, p.~042919, Apr. 2013.

\bibitem{garcia-ojalvo_noise_1999}
J.~Garcia-Ojalvo and J.~Sancho, {\em Noise in Spatially Extended Systems}.
\newblock Institute for Nonlinear Science, Springer, 1999.

\bibitem{blair_clustering_2003}
D.~L. Blair and A.~Kudrolli, ``Clustering transitions in vibrofluidized
  magnetized granular materials,'' {\em Physical Review E}, vol.~67, p.~021302,
  Feb. 2003.

\bibitem{stambaugh_segregation_2004}
J.~Stambaugh, Z.~Smith, E.~Ott, and W.~Losert, ``Segregation in a monolayer of
  magnetic spheres,'' {\em Physical Review E}, vol.~70, p.~031304, Sept. 2004.

\bibitem{miller_stress_1996}
B.~Miller, C.~O'Hern, and R.~P. Behringer, ``Stress fluctuations for
  continuously sheared granular materials,'' {\em Physical Review Letters},
  vol.~77, pp.~3110--3113, Oct. 1996.

\bibitem{seiden_complexity_2011}
G.~Seiden and P.~J. Thomas, ``Complexity, segregation, and pattern formation in
  rotating-drum flows,'' {\em Reviews of Modern Physics}, vol.~83,
  pp.~1323--1365, Nov. 2011.

\bibitem{wassgren_vertical_1996}
C.~R. Wassgren, C.~E. Brennen, and M.~L. Hunt, ``Vertical vibration of a deep
  bed of granular material in a container,'' {\em Journal of Applied
  Mechanics}, 1996.

\bibitem{rosato_perspective_2002}
A.~D. Rosato, D.~L. Blackmore, N.~Zhang, and Y.~Lan, ``A perspective on
  vibration-induced size segregation of granular materials,'' {\em Chemical
  Engineering Science}, vol.~57, pp.~265--275, Jan. 2002.

\bibitem{windows-yule_inelasticity-induced_2014}
C.~R.~K. Windows-Yule and D.~J. Parker, ``Inelasticity-induced segregation: Why
  it matters, when it matters,'' {\em {EPL} (Europhysics Letters)}, vol.~106,
  p.~64003, June 2014.

\bibitem{meerson_close-packed_2003}
B.~Meerson, T.~Pöschel, and Y.~Bromberg, ``Close-packed floating clusters:
  Granular hydrodynamics beyond the freezing point?,'' {\em Physical Review
  Letters}, vol.~91, p.~024301, July 2003.

\bibitem{eshuis_granular_2005}
P.~Eshuis, K.~van~der Weele, D.~van~der Meer, and D.~Lohse, ``Granular
  leidenfrost effect: Experiment and theory of floating particle clusters,''
  {\em Physical Review Letters}, vol.~95, Dec. 2005.

\bibitem{leidenfrost_aquae_1756}
J.~G. Leidenfrost, ``De aquae communis nonnullis qualitatibus tractatus,''
  1756.

\bibitem{eshuis_onset_2010}
P.~Eshuis, D.~van~der Meer, M.~Alam, H.~J. van Gerner, K.~van~der Weele, and
  D.~Lohse, ``Onset of convection in strongly shaken granular matter,'' {\em
  Physical Review Letters}, vol.~104, p.~038001, Jan. 2010.

\bibitem{eshuis_buoyancy_2013}
P.~Eshuis, K.~v.~d. Weele, M.~Alam, H.~J.~v. Gerner, M.~v.~d. Hoef, H.~Kuipers,
  S.~Luding, D.~v.~d. Meer, and D.~Lohse, ``Buoyancy driven convection in
  vertically shaken granular matter: experiment, numerics, and theory,'' {\em
  Granular Matter}, vol.~15, pp.~893--911, Dec. 2013.

\bibitem{coullet_dispersion-induced_1994}
P.~Coullet, T.~Frisch, and G.~Sonnino, ``Dispersion-induced patterns,'' {\em
  Physical Review E}, vol.~49, pp.~2087--2090, Mar. 1994.

\bibitem{clerc_localized_2008}
M.~G. Clerc, S.~Coulibaly, and D.~Laroze, ``Localized states beyond the
  asymptotic parametrically driven amplitude equation,'' {\em Physical Review
  E}, vol.~77, p.~056209, May 2008.

\bibitem{rivas_low-frequency_2013}
N.~Rivas, S.~Luding, and A.~R. Thornton, ``Low-frequency oscillations in narrow
  vibrated granular systems,'' {\em New Journal of Physics}, vol.~15,
  p.~113043, Nov. 2013.

\bibitem{eshuis_phase_2007}
P.~Eshuis, K.~v.~d. Weele, D.~v.~d. Meer, R.~Bos, and D.~Lohse, ``Phase diagram
  of vertically shaken granular matter,'' {\em Physics of Fluids
  (1994-present)}, vol.~19, p.~123301, Dec. 2007.

\bibitem{mcnamara_energy_1998}
S.~McNamara and S.~Luding, ``Energy nonequipartition in systems of inelastic,
  rough spheres,'' {\em Physical Review E}, vol.~58, pp.~2247--2250, Aug. 1998.

\bibitem{goldshtein_mechanics_1995}
A.~Goldshtein, M.~Shapiro, L.~Moldavsky, and M.~Fichman, ``Mechanics of
  collisional motion of granular materials. part 2. wave propagation through
  vibrofluidized granular layers,'' {\em Journal of Fluid Mechanics}, vol.~287,
  pp.~349--382, Mar. 1995.

\bibitem{soto_granular_2004}
R.~Soto, ``Granular systems on a vibrating wall: The kinetic boundary
  condition,'' {\em Physical Review E}, vol.~69, p.~061305, June 2004.

\bibitem{herrmann_modeling_1998}
H.~J. Herrmann and S.~Luding, ``Modeling granular media on the computer,'' {\em
  Continuum Mechanics and Thermodynamics}, vol.~10, pp.~189--231, Aug. 1998.

\bibitem{luding_granular_1995}
S.~Luding, ``Granular materials under vibration: Simulations of rotating
  spheres,'' {\em Physical Review E}, vol.~52, pp.~4442--4457, Oct. 1995.

\bibitem{luding_how_1998}
S.~Luding and S.~McNamara, ``How to handle the inelastic collapse of a
  dissipative hard-sphere gas with the {TC} model,'' {\em cond-mat/9810009},
  Oct. 1998.

\bibitem{windows-yule_thermal_2013}
C.~R.~K. Windows-Yule, N.~Rivas, and D.~J. Parker, ``Thermal convection and
  temperature inhomogeneity in a vibrofluidized granular bed: The influence of
  sidewall dissipation,'' {\em Physical Review Letters}, vol.~111, p.~038001,
  July 2013.

\bibitem{prigogine_self-organization_1977}
I.~Prigogine, {\em Self-Organization in Nonequilibrium Systems: From
  Dissipative Structures to Order through Fluctuations}.
\newblock New York: Wiley, 1 edition~ed., May 1977.

\bibitem{lan_macroscopic_1995}
Y.~Lan and A.~D. Rosato, ``Macroscopic behavior of vibrating beds of smooth
  inelastic spheres,'' {\em Physics of Fluids (1994-present)}, vol.~7,
  pp.~1818--1831, Aug. 1995.

\bibitem{bodenschatz_recent_2000}
E.~Bodenschatz, W.~Pesch, and G.~Ahlers, ``Recent developments in
  rayleigh-bénard convection,'' {\em Annual Review of Fluid Mechanics},
  vol.~32, no.~1, pp.~709--778, 2000.

\bibitem{oh_thermal-noise_2003}
J.~Oh and G.~Ahlers, ``Thermal-noise effect on the transition to
  rayleigh-bénard convection,'' {\em Physical Review Letters}, vol.~91,
  p.~094501, Aug. 2003.

\bibitem{windows-yule_low-frequency_2014}
C.~R.~K. Windows-Yule, N.~Rivas, D.~J. Parker, and A.~R. Thornton,
  ``Low-frequency oscillations and convective phenomena in a density-inverted
  vibrofluidized granular system,'' {\em Physical Review E}, vol.~90,
  p.~062205, Dec. 2014.

\bibitem{luding_simulations_1996}
S.~Luding, E.~Clément, J.~Rajchenbach, and J.~Duran, ``Simulations of pattern
  formation in vibrated granular media,'' {\em {EPL} (Europhysics Letters)},
  vol.~36, p.~247, Nov. 1996.

\bibitem{dawes_emergence_2010}
J.~H.~P. Dawes, ``The emergence of a coherent structure for coherent
  structures: localized states in nonlinear systems,'' {\em Philosophical
  Transactions of the Royal Society A: Mathematical, Physical and Engineering
  Sciences}, vol.~368, pp.~3519--3534, Aug. 2010.

\bibitem{swift_hydrodynamic_1977}
J.~Swift and P.~C. Hohenberg, ``Hydrodynamic fluctuations at the convective
  instability,'' {\em Physical Review A}, vol.~15, pp.~319--328, Jan. 1977.

\bibitem{agez_universal_2008}
G.~Agez, M.~G. Clerc, and E.~Louvergneaux, ``Universal shape law of stochastic
  supercritical bifurcations: Theory and experiments,'' {\em Physical Review
  E}, vol.~77, p.~026218, Feb. 2008.

\end{thebibliography}
\bibliographystyle{ieeetr}

\end{document}